\documentclass[oldversion]{aa}
\usepackage{natbib}
\bibpunct{(}{)}{;}{a}{}{,} 
\usepackage{txfonts}
\usepackage{graphicx}
\usepackage{pstricks}

\newcommand{\add}[1]{{#1}}

\begin{document}
\title{Expansion of magnetic flux concentrations: a comparison of Hinode SOT data and models}
\author{A. Pietarila\inst{1}, R. Cameron\inst{1}, S.K. Solanki\inst{1,2}}
\institute{Max-Planck-Institute for Solar System Research, Max-Planck-Strasse 2, 37191 Katlenburg-Lindau, Germany \and School of Space Research, Kyung Hee University, Yongin, Gyeonggi, 446-701, Korea}
\date{Received <date> / Accepted <date>}

\abstract{{\it Context:} The expansion of network magnetic fields with height is a fundamental property of flux tube models. A rapid expansion is required to form a magnetic canopy.\\
{\it Aims:} We characterize the observed expansion properties of magnetic network elements and compare them with the thin flux tube and sheet approximations, as well as with magnetoconvection simulations. \\
{\it Methods:} We used data from the Hinode SOT NFI NaD$_{1}$ channel and spectropolarimeter to study the appearance of magnetic flux concentrations seen in circular polarization as a function of position on the solar disk. We compared the observations with synthetic observables from models based on the thin flux tube approximation and magnetoconvection simulations with two different upper boundary conditions for the magnetic field (potential and vertical). \\
{\it Results:} The observed circular polarization signal of magnetic flux concentrations changes from unipolar at disk center to bipolar near the limb, which implies an expanding magnetic field. The observed expansion agrees with expansion properties derived from the thin flux sheet and tube approximations. Magnetoconvection simulations with a potential field as the upper boundary condition for the magnetic field also produce bipolar features near the limb while a simulation with a vertical field boundary condition does not. \\
{\it Conclusions:} \add{The near-limb apparent bipolar magnetic features seen in high-resolution Hinode observations can be interpreted using a simple flux sheet or tube model. This lends further support to the idea that magnetic features with vastly varying sizes have similar relative expansion rates. The numerical simulations presented here are less useful in interpreting the expansion since \add{the diagnostics we are interested in are} strongly influenced by the choice of the upper boundary condition for the magnetic field in the purely photospheric simulations.}}

\titlerunning{Expansion of magnetic flux concentrations}
\authorrunning{Pietarila et al}

\maketitle

\section{Introduction}

At present, observations do not provide an explicit picture of how the chromospheric network magnetic field is structured. On one hand, we have increasing observational evidence of
something which can be interpreted loosely as a canopy structure:
e.g., fibrils in the Ca II infrared triplet lines;
\citealt{Vecchio+others2007}, large-scale canopy structures in combined
Zeeman and Hanle studies: \cite{Bianda+others1998,
 Stenflo+others2002}, canopy-like expansion seen in magnetograms near the limb;
\cite{Jones+Giovanelli1983}). On the other hand, \add{we know that 
the Sun is more complicated than implied by simple models \citep{WedemeyerBoehm09}. An important question 
is to what extent simple models of flux tubes are able to reproduce 
the center-to-limb appearance of the network magnetic field structures.}

The appearance of network magnetic flux concentrations in circular
polarization maps changes from unipolar at disk center to bipolar near
the limb. This is consistent with network magnetic fields expanding
and fanning out with height as proposed by \cite{Gabriel1976}. In the chromosphere, the fanning is clearly present; e.g., \cite{Jones+Giovanelli1983} found low-lying,
200-800 km, magnetic canopies in magnetograms taken near the solar
limb. More recently \cite{Kontar+Hannah+MacKinnon2008} has used hard X-ray
observations from RHESSI to estimate the expansion and found that the
magnetic field expanded noticeably at a height of $\approx$ 900 km.

 Expansion of magnetic field with height has been studied in photospheric structures mostly using magnetic flux tube models. Indeed, flux tube models predict a rapid expansion of the field with height. \cite{Solanki+others1999} shows that magnetic
structures as different in size and flux as small flux tubes and
sunspots have similar relative expansion rates, which agree with the thin flux tube approximation. A study of the characteristics of magnetic flux structures in radiative magnetohydrodynamic (MHD) simulations revealed the expansion properties to be similar with the thin flux tube and sheet approximations (\citealt{Yelles+others2009}). The expansion is seen in observations: e.g., a thin flux tube model can simultaneously reproduce the observed Zeeman splittings of Mg 12.32 $\mu$m, Fe 525.0 nm and Fe 1.56 $\mu$m lines, which span the upper to the lower photosphere in formation height \citep{Bruls+Solanki1995}. Bruls \& Solanki also showed that a flux tube model can explain the Mg 12.32 $\mu$m line profile shapes observed by Zirin \& Popp (1989). Additional evidence of expansion is that a canopy resulting from the expansion of a flux tube can best explain the observed photospheric asymmetric Stokes $V$ profiles with weak zero-crossing shifts \citep{Grossmann-Doerth+others1988}.

\nocite{Zirin+Popp1989} 

In this paper we use circular polarization maps from the Solar Optical Telescope (SOT, \citealt{Tsuneta+others2008}) on the Hinode satellite to study the
\add{average} expansion properties by characterizing how the appearance of network
flux concentrations changes from the solar disk center to the
limb. The center-to-limb approach lets us examine the expansion at
different viewing angles and at different heights due to the shift in
 the height \add{range where} spectral lines \add{are formed} as a function of $\mu$
($\mu=cos(\theta)$, where $\theta$ is the viewing angle). To further
expand the coverage we use SOT observations from the
spectropolarimeter (SP) and the narrowband filter imager
(NFI) NaD$_{1}$ channel. We combine the observations with
modeling the expansion of magnetic flux with height \add{at various $\mu$-values} by using the thin
flux tube and sheet approximations and more realistic 3-dimensional magneto-convection
simulations.

\section{Observations and models}

\subsection{Observations and analysis}

The SOT NFI NaD$_{1}$ (589.6 nm, effective $g$ factor 1.33) data used in this study consist of 19 circular polarization
filtergrams of non-active region maps at different positions on the
solar disk (see Table~\ref{tab:nfi-sp-dat} for time, location, exposure time and field of view, where $xcen$ and $ycen$ are orthogonal distances from Sun center and $xcen$ is directed along the equator). The data reduction was done
using the Solar Software (SSW) \footnote{http://www.lmsal.com/solarsoft} package routine {\it fg\_prep.pro}. To increase the signal-to-noise ratio, frames taken
within 3 minutes are summed together (number of frames is shown as a multiplication factor in from of the exposure time in Table~\ref{tab:nfi-sp-dat}). Errors in pointing are corrected
in the limb data sets by forcing $\mu$=0 to coincide with the visible
solar limb. The NaD$_{1}$ line is often called a chromospheric diagnostic but 3-dimensional non-local thermodynamic equilibrium (nonLTE) radiative transfer calculations of magnetoconvection simulations have demonstrated that the line is to a large degree photospheric, especially in magnetic flux concentrations \citep{Leenaarts+others2009}.

\begin{table*}
\caption{NFI and SP data sets}
\label{tab:nfi-sp-dat}
\centering
\begin{tabular}{c c c c c c c}
\hline \hline
NFI & & & & & & \\
 & date & xcen [$^{\prime \prime}$] & ycen [$^{\prime \prime}$] & $\mu$ & exp [s] & FOV [$^{\prime \prime}$ $\times$ $^{\prime \prime}$]\\
\hline \hline
       1& 2008-01-16T11:13:34.821 &           4 &           0 &1.00& 4
 $ \times $ 0.20& 113. $  \times  $ 113.\\
       2& 2008-04-03T12:55:37.304 &         100 &        -270 &0.95& 4
 $ \times $ 0.20& 64.0 $  \times  $ 164.\\
       3& 2007-11-11T13:07:55.041 &        -313 &         -57 &0.94& 10
 $ \times $ 0.20& 41.0 $  \times  $ 164.\\
       4& 2007-12-01T11:04:34.300 &         325 &           3 &0.94& 4
 $ \times $ 0.12& 225. $  \times  $ 113.\\
       5& 2007-11-29T11:28:37.095 &         377 &        -184 &0.90& 4
 $ \times $ 0.20& 113. $  \times  $ 113.\\
       6& 2007-11-05T05:28:45.378 &         503 &         109 &0.84& 10
 $ \times $ 0.20& 41.0 $  \times  $ 164.\\
       7& 2007-11-05T07:27:33.212 &         518 &         110 &0.83& 10
 $ \times $ 0.20& 41.0 $  \times  $ 164.\\
       8& 2008-05-13T16:27:02.821 &         511 &         187 &0.82& 7
 $ \times $ 0.20& 56.3 $  \times  $ 113.\\
      9& 2007-11-15T10:09:35.952 &         -58 &        -561 &0.81& 10
 $ \times $ 0.20& 41.0 $  \times  $ 164.\\
      10& 2007-11-15T08:23:31.692 &         -71 &        -561 &0.81& 10
 $ \times $ 0.20& 41.0 $  \times  $ 164.\\
      11& 2008-04-05T20:34:34.811 &         542 &        -291 &0.77& 4
 $ \times $ 0.20& 64.0 $  \times  $ 164.\\
      12& 2007-11-13T00:44:33.710 &        -303 &        -551 &0.76& 10
 $ \times $ 0.20& 41.0 $  \times  $ 164.\\
      13& 2008-05-14T12:51:32.612 &         735 &         -60 &0.64& 4
 $ \times $ 0.20& 64.0 $  \times  $ 164.\\
      14& 2007-11-17T10:45:35.227 &         858 &        -114 &0.43& 3
 $ \times $ 0.20& 164. $  \times  $ 164.\\
      15& 2008-03-24T14:28:38.464 &         -10 &        -947 &0.31& 4
 $ \times $ 0.20& 64.0 $  \times  $ 164.\\
      16& 2008-03-22T16:48:40.335 &         -10 &        -946 &0.31& 4
 $ \times $ 0.20& 64.0 $  \times  $ 164.\\
      17& 2007-11-10T11:06:02.179 &          -6 &         928 &0.25& 7
 $ \times $ 0.20& 164. $  \times  $ 164.\\
      18& 2007-11-12T11:37:31.448 &         230 &         912 &0.20& 7
 $ \times $ 0.20& 164. $  \times  $ 164.\\
      19& 2007-12-19T09:34:40.109 &         931 &        -163 &0.19& 4
 $ \times $ 0.15& 113. $  \times  $ 113.\\

\hline
SP & & & & & \\
\hline
1& 2007-09-06T15:29:04.938 &         -46 &           6 &1.00&8.0 & 149 $\times$ 162 \\
2& 2007-09-10T15:03:12.528 &          87 &           7 &1.00&9.6 & 56  $\times$  162 \\
3& 2007-09-22T09:28:05.954 &          27 &         307 &0.95&13. & 47  $\times$  162\\
4& 2007-09-21T11:28:34.622 &          25 &         607 &0.77&13. & 47  $\times$  162\\
5& 2007-09-10T22:13:23.366 &        -247 &         656 &0.68&1.6 & 153  $\times$  162\\
6& 2008-01-21T06:02:48.747 &         760 &        -387 &0.46&13. & 56  $\times$  162\\
7& 2007-09-07T03:59:46.805 &         -18 &         922 &0.28&9.6 & 105  $\times$  162\\
8& 2007-09-03T16:06:00.480 &         144 &         922 &0.23&9.6 & 298  $\times$  162\\
9& 2007-11-03T05:17:25.426 &          54 &         952 &0.11&13. & 208  $\times$  162\\
10& 2007-11-01T06:18:20.542 &          93 &         952 &0.080&13. & 248  $\times$  162\\

\hline
\end{tabular}
\end{table*}

For SP data we use 10 raster scans with exposure times of eight seconds or more (i.e., deep magnetogram mode) at varying positions on the
solar disk (see Table~\ref{tab:nfi-sp-dat}) reduced using the Solar
Software package routines {\it sp\_prep.pro} and {\it stksimages\_sbsp.pro}. The step size is 0.149 $^{\prime \prime}$ in all scans but number 5 where it is 0.29 $^{\prime \prime}$. The reduced data consist of maps of circular polarization and
linear polarization in the 630.1 (effective $g$=1.67) and 630.2 nm (effective $g$=2.5) Fe I lines, both combined and
separate. For details of the routines see \cite{Lites+others2008}. As in NaD$_{1}$ data, pointing errors are corrected in maps where the solar limb is visible.

Flux concentrations (coherent features with circular polarization
signal clearly above the noise level) are identified manually in each
NFI and SP circular polarization image. For the identification we use images of the absolute value of the flux instead of the signed flux to reduce the bias towards choosing bipolar features near the limb. A central pixel of each identified feature is chosen and a vector connecting the disk center and limb passing
through this point is defined. This vector is the \add{average} radial cut of the
feature. The SP circular and linear polarization radial cuts are made by averaging over 7 pixels
perpendicular to the vector for the 630.1 and 630.2 nm. Unless otherwise specified we use the average of the two lines. For NFI data the radial cuts (width 7 pixels) are
made of the Stokes $V$ filter signal. No correction for foreshortening
is made. Finally, peaks (location and amplitude) in the radial cuts are
identified manually in each cut (see Fig.~\ref{fig:nad-e1}). \add{Since we are interested in the large scale expansion of the field as opposed to a detailed study on the small scale morphology of the structures, spatial and, in the case of NaD$_{1}$ also temporal averaging, are justified.}  
  
\subsection{Models}

We have employed two independent sets of models, the simple thin-tube approximation and realistic 3-D simulations. 
\add{The thin flux tube model is based on the assumption that the flux tube is much thinner than the pressure and density scale heights. The plasma and magnetic properties inside
the tube can then be expanded in terms of a low order polynomial \citep{Defouw1976, Roberts+Webb1978}.  
Here we use the lowest order polynomial possible:} $0^{th}$ order in radius, $r$, for the vertical component
of the magnetic field, $B_{z}$, and $1^{st}$ order in the radial component
of the field, $B_{r}$. 
\add{The component of the 
Lorentz force corresponding to magnetic tension does not appear in this $0^{th}$ order in $r$ expansion of $B_{z}$ (it first appears in the $2^{nd}$ order expansion). We use
semi-empirical atmospheric models for the internal and external models, from which the magnetic field
can be derived using the condition that the total (gas + magnetic) pressure inside the
tube at each height is equal to that of the external gas pressure.}

The external non-magnetic atmosphere is given by the semi-empirical SRPM
model of the quiet Sun low chromosphere \citep{SRPM2007} and internal
atmosphere by the recent facular and plage models of the photosphere
\citep{newFAL2006} which have been extended in height by adding the
FALF (facular model, \citealt{FAL1993}) model chromosphere. \add{The effect of internetwork fields is not considered in the external model. This is justified since their strength in the lower photosphere is on the order of a 100 G, i.e., the magnetic pressure is 
roughly 1\% of the gas pressure. At greater heights these internetwork fields fall off rapidly \citep{Schuessler08}
and so remain unimportant to this particular study. }
We produce 2-dimensional atmospheric slabs ($x,z$) by varying the flux tube/sheet parameters:
radius at $z$=0 km ($r_{0}$), field strength at $z$=0 km ($B_{0}$), and
the internal flux tube/sheet atmosphere (facular or plage). 

Observables are synthesized under the assumption of LTE using the Nicole-code of Socas-Navarro \citep{Melanie}. Arguably, especially the NaD$_{1}$ line is influenced by nonLTE effects, but since we consider the Stokes $V$ signal which has its' maximum value away from the line core \add{(where nonLTE effects are significant)} and , more specifically, in the ratio of the Stokes $V$ amplitudes, we chose to do the synthesis in LTE. We are not interested in comparing absolute values of the amplitudes or details of the line profiles, which may well be affected by nonLTE effects, but rather in the ratio between the amplitudes of the center- and limb-sides of magnetic concentrations, $\frac{a_c}{a_l}$, which should be independent of nonLTE effects, at least to first order. We compute 1-dimensional rays
passing through the slabs at various line of sight (LOS) angles and produce synthetic observables. Synthetic cuts are
created from the Stokes $V$ amplitudes and areas (sign determined by the sign of the blue lobe). This is done also for the NaD$_1$ line instead of using the filter transmission profile. By doing this the NaD$_1$ synthetic observables pick up the maximum signal at each point. We compute the observables\add{, radial cuts of Stokes $V$ areas and amplitudes,} using the original tube/sheet model resolution (10 km in the horizontal direction, rays placed at 10 km intervals) and also for the spectral (90 m\AA\ for NFI, 23 m\AA\ for SP) and spatial (theoretical point spread function (PSF) for a 50 cm telescope, 0.08$^{\prime \prime}$ and 0.16$^{\prime \prime}$ pixels size for NFI and SP, respectively) resolutions comparable to those of Hinode. \add{For the purposes of this study, we ignore contributions from the spider and other optical elements of Hinode on the PSF. Additionally, we assume the PSF to be 1-dimensional. These idealizations are appropriate since we are not interested in the detailed structure of the features in our simplified model.}

For the purposes of this paper the difference between flux tubes and sheets lies in their linear expansion rates. In the thin tube approximation the expansion is driven by the magnetic flux conservation with height: $BA=const$, where $B$ is the field strength and $A$ is the tube's cross-section. As $B$ drops roughly exponentially with height, $A$ increases inversely. In a flux tube the increase in radius goes as $r \sim \sqrt{A}$, while in a sheet the magnetic field only expands in the direction perpendicular to the sheet, so that $r \sim A$. Consequently, $r$ increases far more rapidly in sheets than tubes. \add{The ratio, $\frac{a_c}{a_l}$, strongly depends on the viewing angle. It is smallest when rays are perpendicular to the sheet direction and infinite, i.e., unipolar, when rays are parallel. Therefore, to delimit the range} we only synthesize radial cuts for rays pointing in the perpendicular direction. The expansion rate for a magnetic feature consisting of several flux tubes, which merge above a height \add{determined by the density of the features and their magnetic flux} follows the same scaling. Below the merging height the flux tubes expand individually. Above this, they expand as a single feature following $A(z) \sim A_{0}/B(z)$, where $A_{0}$ is the summed cross-section of the individual tubes at $z=0$.


To study flux expansion under more realistic conditions we use
radiative magnetohydrodynamic (MHD) simulations. \add{In contrast to the simple flux tube model, the simulations allow us to look in detail at the effect of small scale, inhomogeneous structures on the expansion. We use} the MuRAM
code (MPS/University of Chicago Radiative MHD,
\citealt{Voegler+others2005}). The code takes into account effects of
full compressibility, open boundary conditions as well as non-gray
radiative transfer and partial ionization. 
In the current study the
simulation domain is 24Mm$\times$24Mm$\times$1.68Mm
(576$\times$576$\times$120 grid points) of which approximately 700 km
is above the $log(\tau_{500nm})=0$ level. 
\add{To construct our initial condition we began with a snapshot
from a non-magnetic convection run. To this we added a magnetic 
strip with a squared Gaussian profile. We chose a peak value of
300 G so that the initial magnetic field is energetically important
above the surface but much less so below the surface. The full width
at half maximum was chosen to be 3.157 Mm, so that the patch of 
magnetic field spans several granules. This setup is reminiscent of
an isolated, unipolar network lane. The field then evolves on several 
timescales. The slowest of these is the dispersion of magnetic flux 
due to the granular motions advecting them. This process can be
modeled as a random walk with a length-scale of 
approximately 1~Mm and a timescale of approximately 5-10 minutes. A second important timescale is that required to achieve pressure balance
in the vertical direction and for our box, which extends $700$~km
above approximately $log(\tau_{500nm})=0$ surface and where the sound-speed 
is $\approx 10$~km/s, is some small multiple of 700~km/Cs=70 s. 
The time-scale for pressure balance to be achieved in the horizontal
direction should be similar because for fields of substantial strength,
the length scales of the fields in the horizontal and vertical
directions are likely to be similar. An intermediate timescale is
that of the magnetic flux to be expelled from the interior of the
granules, which takes place on the convective overturning timescale of
the flows. Since we study the expansion with height
of a thin strip of magnetic field, we allowed the simulation to evolve
from the initial condition for a few granular lifetimes ($\approx$ 15 minutes).
The time-development of the magnetic field after this time will be the
subject of a planned future paper. }

We make two sets of simulations with two
different upper boundary conditions (BCs) for the magnetic field:
vertical and potential field. \add{In the potential filed case we considered both a case where the field was vertical at infinity and where it was slightly inclined. The inclination was found to make a negligible difference for the expansion}. Otherwise the BCs for the two
simulations are the same: velocity BC is open at the top and bottom,
and all BCs are periodic in the horizontal directions. As for the thin
flux tube slabs we produce synthetic observables at varying LOS angles
from snapshots of the simulation taken after the main
transient phase due to the initial conditions is over 
(approximately 12 minutes in solar time). Since the simulations
extend only to approximately 700 km above the $\tau$=1 surface, the
observables are formed in a region strongly influenced by the
BCs \add{(this is discussed in section 4.2)}. Rays entering at an inclined LOS are even more strongly influenced by this. These
caveats as well as the assumption of LTE for the radiative transfer
should be kept in mind when comparing with observations.
  
\section{Hinode observations}
\subsection{ NaD$_{1}$ filtergrams}

\begin{figure}
\includegraphics[width=88mm]{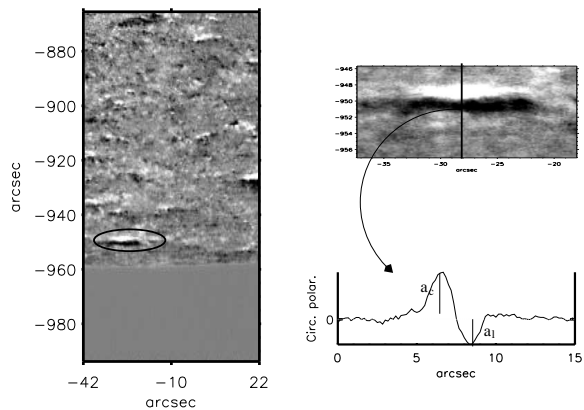}
\caption{Example of radial cut in NaD$_{1}$ filtergrams near the limb. The
  image on left shows the full field of view with the magnetic feature
  of interest circled. On top right the smaller image is a zoom in of
  the feature and the plot below shows the radial cut. The center-side
  and limb-side amplitudes, $a_c$ and $a_l$ are indicated.}
\label{fig:nad-e1}
\end{figure}

Due to expansion of magnetic flux with height the
appearance of magnetic features in NaD$_{1}$ \add{Stokes $V$} filtergrams changes from
unipolar at disk center to bipolar close to the limb
(Fig.~\ref{fig:nad-e1}). Additionally, due to foreshortening effects
the features become more elongated parallel to the limb. The size of the apparent bipolar features varies in the direction
parallel to the limb from a few arcseconds to well over
ten. In the radial direction the size (without correcting for
foreshortening) is around four arcseconds at $\mu < 0.4$. No
significant further change in size is seen towards the limb. As
one moves away from the disk center towards the limb the circular polarization radial cuts change
from having a single peak (unipolar cuts) to having two peaks with opposite signs (bipolar cuts). The
limb-side peak is usually smaller. The ratio of the peaks ($\frac{|a_{cen}|}{|a_{limb}|}$, see
Fig.~\ref{fig:nad-e1}) decreases toward the limb and will be used as a proxy for expansion.

\begin{figure}
\includegraphics[width=8.8cm]{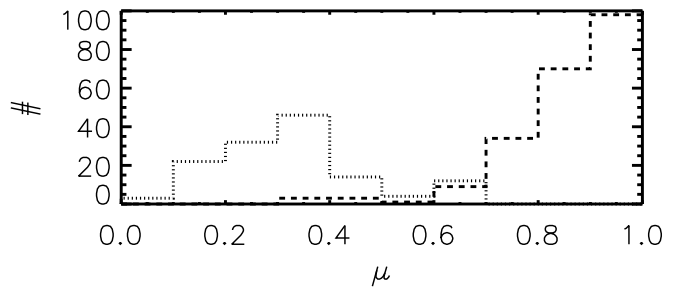}
\caption{Histogram of number of uni- (dashed line) and bipolar (dotted
  line) circular polarization radial cuts in the NFI data as a
  function of $\mu$.  }\label{fig:nad-hist}
\end{figure}

 Histograms of the number of unipolar and bipolar radial cuts as a function of $\mu$ (Fig.~\ref{fig:nad-hist})
 display the change from unipolar to bipolar: at disk center nearly
 all cuts are unipolar with either one or two peaks. The
 double-peaked cuts are from features with complex, i.e., not circular
 or compact, shapes. At the limb all cuts are double-peaked
 with oppositely signed peaks, i.e., bipolar. The apparent bipolar features
 become dominant at $\mu < 0.5$. Because of the insufficient sample it is not possible to say what the statistics are in $\mu$-range 0.5-0.7 (very few observations at those $\mu$-values).

\begin{figure}
\includegraphics[width=8.8cm]{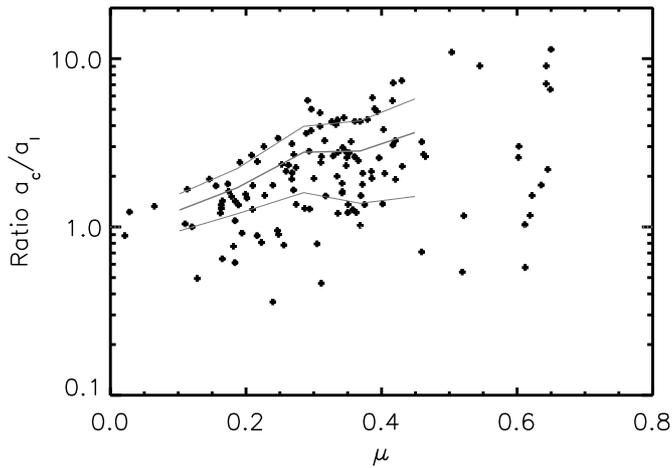}
\caption{Scatter plot of NaD$_{1}$ ratios as a function of $\mu$. Thick solid line shows the mean
  ratio in 0.1 $\mu$ bins and the thin lines are the mean plus/minus
  standard deviation.  }\label{fig:nad-ratio}
\end{figure}

In most apparent bipolar features (102 of 115) the center-side peak has a larger amplitude than the limb-side peak (Fig.~\ref{fig:nad-ratio}). The ratio of the center- to the limb-side peak amplitude
strongly decreases towards the limb. \add{Some scatter in Fig.~\ref{fig:nad-ratio} could be due to the network magnetic fields being inclined in the direction of the LOS or away from it. However, observations of photospheric network magnetic fields indicate that the field inclination is small (less than 10 deg, \citealt{SanchezAlmeida+MartinezPillet1994}) making inclined fields an unlikely explanation for most of the observed scatter.}  No
difference in the ratio is seen for features at the polar limb and at
the equatorial limb, i.e., no significant difference is seen when
using the ratio as a proxy for expansion in the mostly unipolar polar regions
and the likely more heterogeneous equatorial limb. This suggests that the expansion studied here is driven locally by the properties of the individual magnetic features, independent of the magnetic surroundings. \add{The difference between quiet Sun and coronal holes is most pronounced in the upper layers of the atmosphere. Since the observables are formed low in the atmosphere, it is not surprising that no difference is seen with the current expansion diagnostic. Wiegelmann \& Solanki (2004) showed using magnetic field extrapolations that in the quiet Sun both small and large loops are present while in coronal holes the large loops are replaced by open field lines. The far more numerous small loops are more likely to have an effect on the expansion in the photosphere than the large loops.}
        
\nocite{Wiegelmann+Solanki2004}

\subsection{SP scans}

\begin{figure}
\includegraphics[width=8.8cm]{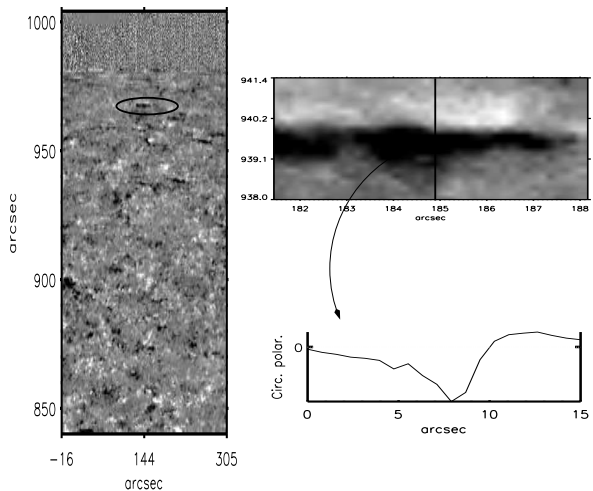}
\caption{As Fig.~\ref{fig:nad-e1} but for the averaged circular polarization signal in the 630 nm Fe lines. 
}\label{fig:fe-e1}
\end{figure}

\begin{figure}
\includegraphics[width=8.8cm]{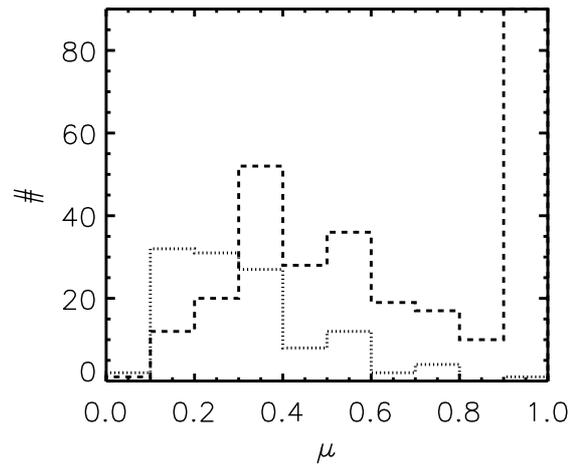}
\caption{Histogram of number of uni- (dashed line) and bipolar (dotted line) cuts as a
  function of $\mu$ for the SP data.}\label{fig:fe-hist}
\end{figure}

Bipolar features are also seen in SP circular polarization maps close to the limb (Fig.~\ref{fig:fe-e1}). However, unlike the NaD$_{1}$ filtergrams, in the purely photospheric iron
lines a significant number of single-peaked cuts are found even at small $\mu$-values (Fig.~\ref{fig:fe-hist}). Furthermore, the flux concentrations close to the limb often look more diffuse in Fe than in NaD$_{1}$ making it difficult to judge where one feature ends and another one begins. Apparent bipolar features outnumber unipolar features at $\mu < 0.3$.

As in NaD$_{1}$ the ratio of the center-side to the limb-side peak amplitude decreases towards the limb, although not so clearly as for NaD$_{1}$. Also, the ratio in Fe is
roughly a factor two larger than in NaD$_{1}$. A direct comparison between the two is not strictly valid since the parameters used for measuring the circular polarization are not the same (filtergram data with a specific filter transmission profile in NaD$_{1}$ and integrated area of circular polarization over the Stokes $V$ spectral
profile for the Fe lines) nor do the measurements have the same signal-to-noise ratio. Additionally, the line formation characteristics, such as Land\'e $g$-factors, are different. The Fe lines likely also sample weaker magnetic fields not visible in the NaD$_{1}$ NFI observables. 

\begin{figure}
\includegraphics[width=8.8cm]{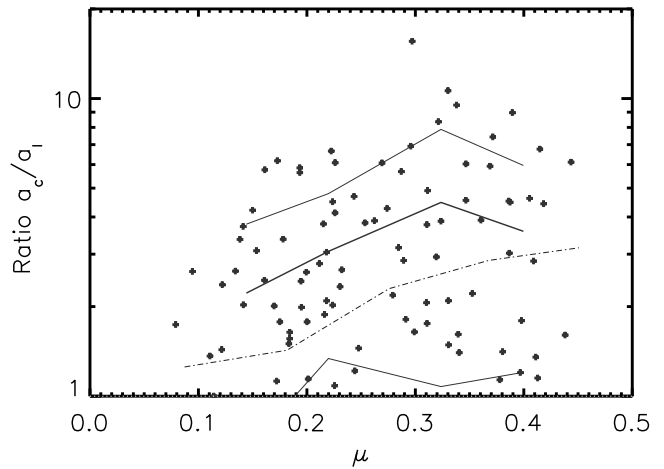}
\caption{Scatter plot of the SP data ratios as a function of $\mu$. Thick solid lines show the mean
  ratio in 0.1 $\mu$ bins and the thin lines are the mean plus/minus
  standard deviation. The
  dash-dotted line is the mean (in 0.1 $\mu$ bins) of the NaD$_{1}$
  ratio. }\label{fig:fe-ratio}
\end{figure}

The peak amplitude of unipolar features' radial cuts decrease
towards the limb although the scatter is very large. At disk center ($\mu >
0.8$) in 48 of 63 cases the circular polarization signal is larger in
630.1 than 630.2 indicating the 630.2 line may be Zeeman
saturated. For off-disk center cuts ($\mu < 0.8$)
only 3 out of 66 cases have a larger peak amplitude in 630.1 than
630.2. The decrease in peak amplitude of unipolar features towards the
limb is due to the line-of sight becoming less aligned with the axis of the magnetic flux tubes (assuming that the magnetic field in the larger magnetic flux
concentrations sampled in this study is mainly vertical in the
photosphere) and the increasing formation height combined with the
decrease in field strength with height \citep{Solanki+others1996}. The peak amplitude of unipolar cuts as a function of $\mu$ is
similar to the center-side amplitude in the bipolar cuts. The total
polarization (circular and linear) is larger in the bipolar
features consistent with the unipolar features being smaller
flux concentrations which do not expand as strongly as the bipolar
features. Since we do not see as many unipolar features in NaD$_{1}$ (in fact none were included in the analysis) this
indicates that either the unipolar features further expand with height (assuming the NaD$_{1}$ signal is formed higher than the Fe lines and appear bipolar in NaD$_{1}$, or they are not strong enough to reach or to be visible
at the formation height of the NaD$_{1}$ signal in the NFI data. Due to the larger width and lower Land\'e $g$-factor of the NaD$_{1}$ line, together with the NFI not being as sensitive as SP, the NFI data may sample only the stronger, larger magnetic features. Note that the NaD$_{1}$ signal does not necessarily \add{always} come from a greater height \citep{deWijn+others2009}.

\section{Synthetic data}
\subsection{Thin flux tube and sheet approximations}

\begin{figure}
\includegraphics[width=8.8cm]{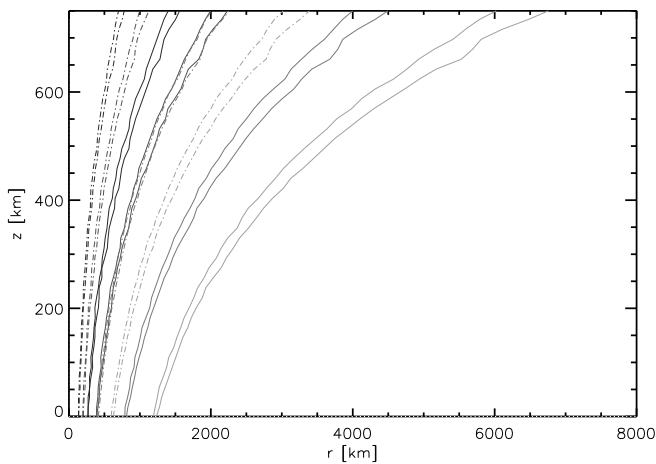}
\caption{Radius of thin flux tubes as a function of height. Colors
  correspond from darkest to lightest to 140, 200, 400 and 600 km
  $r_0$. Thick line is for $B_0$=1300 G and thin line for 1400
  G. Shown are also facular (solid line) and plage (dash-dotted line)
  models as internal atmospheres, but the difference between the two
  is barely visible. }\label{fig:tube-r}
\end{figure}

Besides the geometry (tube or sheet) the choice of flux tube radius at $z$=0 km, $r_0$, is the most important factor
determining the expansion of the thin flux tubes (Fig.~\ref{fig:tube-r}). Note
that the values of $r_0$ considered here (140, 200, 400 and 600 km. $z=0$ corresponds to $\tau=1$ in the external atmosphere) are comparable to or larger than the
pressure scale height, i.e., outside the strict validity range of
the thin flux tube approximation. The absolute expansion is
significantly stronger for $r_0$=400 and 600 km than for the tubes with smaller radii (although the relative expansion, $r(z)/r(0)$, is the same for all). Compared to the choice of radius, the choice of internal model (facular or plage) or $B_0$ (magnetic field strength at $z$=0 km, 1300 or 1400
G) does not produce dramatically different expansion rates as shown in Fig.~\ref{fig:tube-r}. The plage and network models used here describe the network and plage regions as a whole, i.e., they do not specifically address flux tubes. The discrepancy between these models and flux tube models, such as, e.g., \cite{Briand+Solanki1995}, is largest in the low photosphere where the temperature rise is stronger in the flux tube models. This leads to slightly stronger expansion of the tube in the lower layers. In the following, we show synthetic observables from slabs with the facular model
as the internal atmosphere and $B_0$=1300 G, where $B_{0}$ corresponds to $B$ in the the flux tube at $=0$. Ratios resulting from different models (atmospheric model, $B_0$) are not significantly different from the ones shown here. 

\begin{figure}
\includegraphics[width=8.8cm]{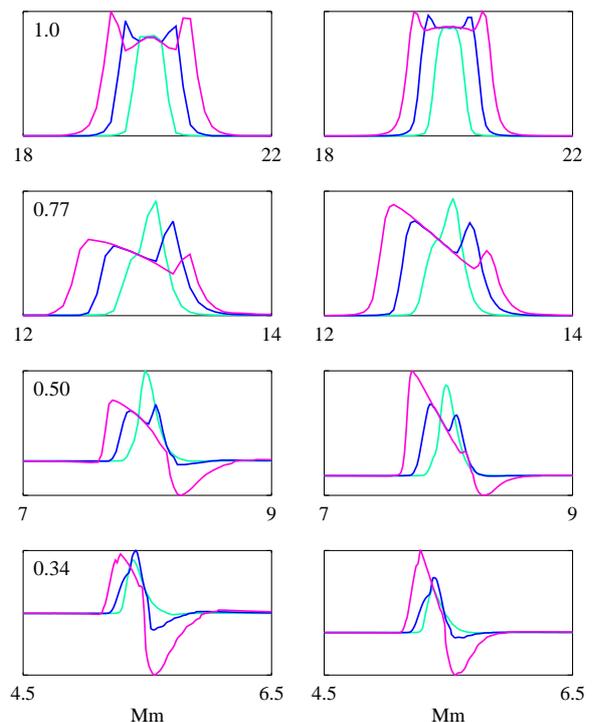}
\caption{Synthetic radial cuts (Hinode resolution) from thin flux tubes at different $\mu$, decreasing from top to bottom ($\mu$ is given within each of the frames in the left column) The various curves refer to different $r_0$: 200 km in turquoise, 400 km in blue, and 600 km in pink. In left column the averaged Fe 620 nm and in the right column Na D$_1$ Stokes $V$ amplitudes (sign given by sign of blue lobe) are plotted.  }\label{fig:tube-res}
\end{figure}

\begin{figure}
\includegraphics[width=8.8cm]{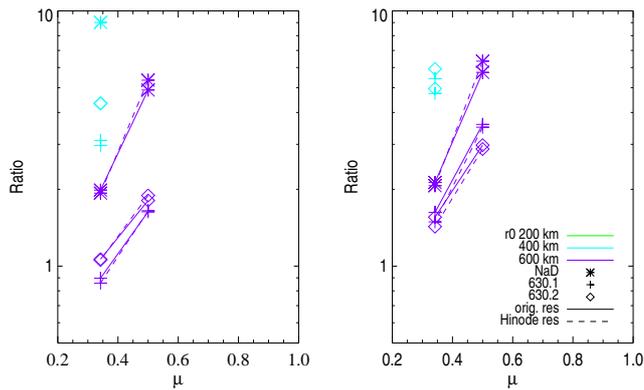}
\caption{Amplitude (left) and area (right) ratios from synthetic radial cuts produced from the thin flux tube models. Shown are only ratios smaller than ten. The amplitudes and areas are computed from Stokes $V$ profiles which have not been normalized to the continuum intensity. }\label{fig:tube-ratio}
\end{figure}

 The radial cuts in NaD$_{1}$ and Fe
become more asymmetric towards the limb (Fig.~\ref{fig:tube-res}). Local maxima in the cuts are
produced close to the boundaries of the flux tube due to the hot-wall effect. When the tube radius, $r_{0}$, is comparable to the pressure scale
height (140 km or 200 km) the tube does not expand enough or at
heights low enough to produce clearly bipolar features in the
Fe and NaD$_{1}$ observables as shown in Fig.~\ref{fig:tube-ratio}\add{, where the results are summarized for all tubes with ratios below ten}. Only
when the tube radius is $\ge$ 400 km do bipolar features begin to appear at
$\mu$=0.5. The smallest ratios are from the $R_{0}$=600 km case where the expansion of the magnetic field is the strongest. Using the amplitude instead of area results in slightly different ratios. The differences between area and amplitude are largest for the smallest $\mu$-values, but both observables show the same trends. The effect of convolving and rebinning the observables to Hinode resolution only slightly alters the ratios. Normalizing the Stokes $V$ profiles to the local continuum intensity prior to computing the observables (area, amplitude) slightly changes the ratios. The NaD$_{1}$ ratios are slightly higher than the Fe ratios, contrary to the observations. For all the observables (original resolution, Hinode resolution, area, amplitude, normalized or not) the trend is the same: tubes with $r_{0}$=400 or 600 km produce bipolar features at $\mu$-values below 0.5 and the ratios decrease towards the limb. The values of the resulting ratios are consistent with the observed ones.

The radial extent of the apparent bipolar features in the flux tubes is at
most $\approx$3 arcseconds at disk center. The sizes are even smaller close to the limb. No large differences are seen in width when comparing the synthetic Fe and NaD$_{1}$ cuts, but in general the features appear wider in Stokes $V$ area cuts than amplitude cuts. This is due to the profiles near the flux tube boundaries having asymmetric, multi-lobed shapes. The integrated absolute area in these profiles can be relatively large whereas the amplitude tends to be fairly small. If the foreshortening effect close to the limb
is taken into account the synthetic features are significantly smaller than
those observed. In the disk center NFI data, diameters of network flux
concentrations range on average from 1 to 3 Mm, i.e., the $r_{0}$=600
km tube is still relatively small. Such small features as these resulting from the flux tube models would be very
difficult to identify close to the limb. An important factor not included in the present analysis is that we expect the network to be composed of many smaller flux tubes/sheets, which may not expand like a single sheet/tube. We address this with the MHD simulations (next section).  

\begin{figure}
\includegraphics[width=8.8cm]{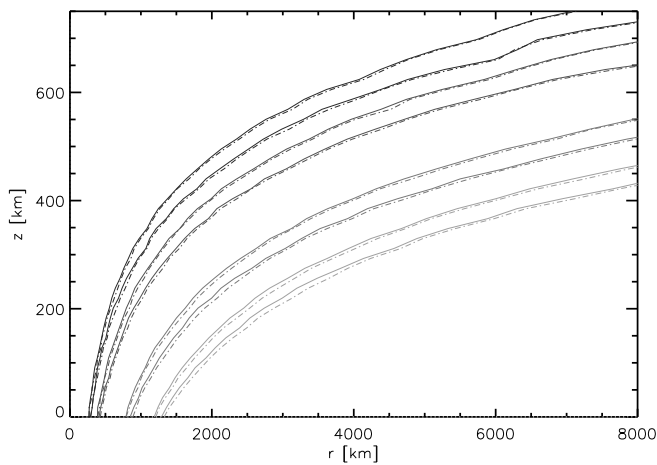}
\caption{As Fig.~\ref{fig:tube-r} but for the thin flux sheet models.}\label{fig:sheet-r}
\end{figure}

\begin{figure}
\includegraphics[width=8.8cm]{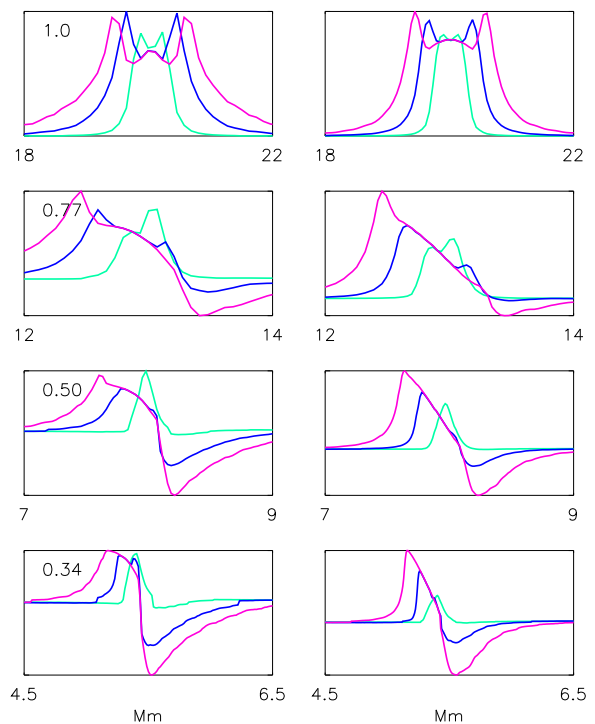}
\caption{As Fig.~\ref{fig:tube-res} but for the thin flux sheet models. }\label{fig:sheet-res}
\end{figure}

\begin{figure}
\includegraphics[width=8.8cm]{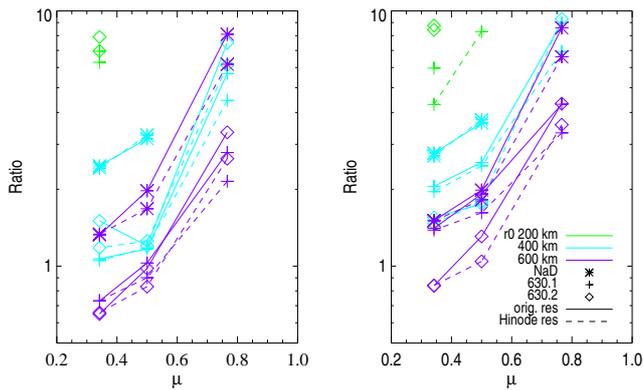}
\caption{As Fig.~\ref{fig:tube-ratio} but for the thin flux sheet models. }\label{fig:sheet-ratio}
\end{figure}

Fig.~\ref{fig:sheet-r} shows the radius of the thin flux sheets as a function of height. As expected, the fanning out is stronger and takes place at lower heights in the sheets than in the tubes. Unlike the tubes, sheets with $r_0$ comparable to the pressure scale height show significant expansion with height. The resulting observables (Fig.~\ref{fig:sheet-res}) differ from the tube case in that 1) the radial cuts become bipolar in appearance at larger $\mu$-values 2) sheets with smaller $r_0$ produce bipolar cuts. Also the radial extent of the apparent bipolar features is larger in the sheets. The ratios (Fig.~\ref{fig:sheet-ratio}) are similar to the tube case. The largest sheets produce bipolar features already at $\mu$=0.77. The switch over from uni- to bipolar in the SOT observations takes place between $\mu$=0.7 and 0.5. As in the tube case the NaD$_{1}$ ratios are slightly larger. Contribution and response functions for the synthetic profiles show that the Stokes $V$ signals of the three lines are formed at a very similar height: the NaD$_{1}$ core is formed higher than the Fe line cores but the Stokes $V$ signal is mostly from the line flanks which are, at least in LTE, mostly formed in the same height range. The only exception are rays entering the flux tube at an inclined angle: compared to the Stokes $V$ signal of the Fe lines, the bulk of the NaD$_{1}$ Stokes $V$ signal from the center-side of the tube is formed slightly higher, where also the \add{line of sight magnetic field} is slightly higher. An additional factor which may play a minor role is saturation. Since the Fe lines are more Zeeman sensitive they become Zeeman saturated at lower field strengths than NaD$_{1}$. Zeeman saturation of profiles from the disk center side of the flux tube would lead to a smaller ratio. We did not explore this further nor did we test varying the microturbulent velocity in the models. Increasing it would broaden the lines and result in a reduced Stokes $V$ amplitude.

\subsection{MuRAM simulations}

\begin{figure}
\includegraphics[width=8.8cm]{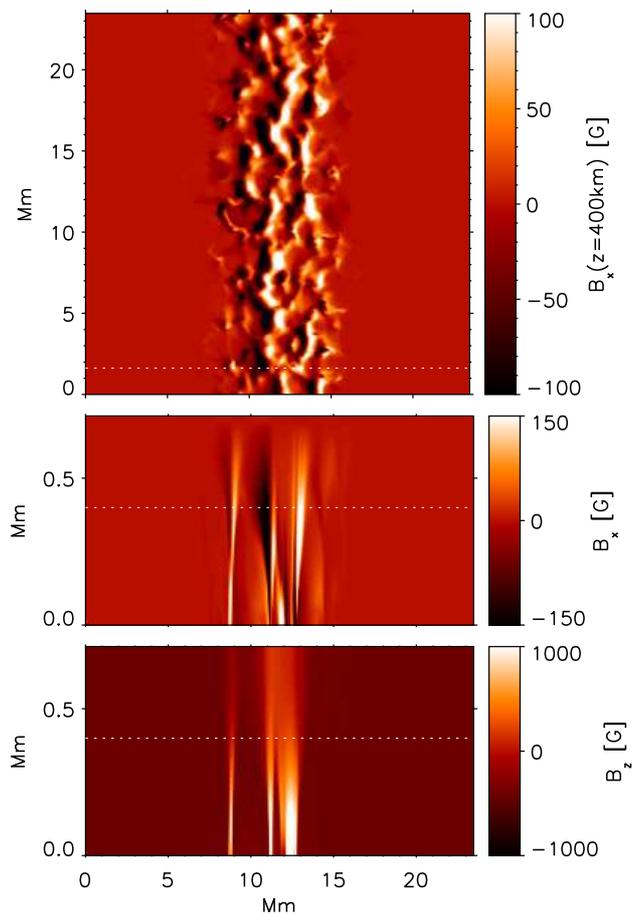}
\caption{Snapshots of the magnetic field in the simulation with a vertical field at the top boundary. Top: horizontal cut through the simulation domain showing the x-component of the magnetic field at a geometric height of roughly 400 km above the $\tau=1$ level. Dotted line shows the location of the vertical cuts shown in the lower frames. Middle and bottom: vertical cuts of the x- and z-component of the magnetic field. Dotted line shows the height of the horizontal cut shown in the top panel. }\label{fig:vert-sim}
\end{figure}

The magnetic field is highly
inhomogeneous in the MuRAM simulations. In the vertical field BC simulation
(Fig.~\ref{fig:vert-sim}) the magnetic field remains confined to the
strip in the middle of the simulation domain also after the transient
phase. Even high up
in the simulation domain the area around the magnetic strip remains void of magnetic field. The main magnetic structure is composed of distinct individual concentrations rooted in the intergranular lanes. These structures do not entirely merge at any height, but their expansion leads to a decrease in field strength with height and only horizontal fields in the individual concentrations have a bipolar-like structure: a horizontal black and white pattern in the horizontal field, i.e., oppositely directed, in Fig.~\ref{fig:vert-sim} is visible only in the individual magnetic flux concentrations and not on a larger spatial scale. 

\begin{figure}
\includegraphics[width=8.8cm]{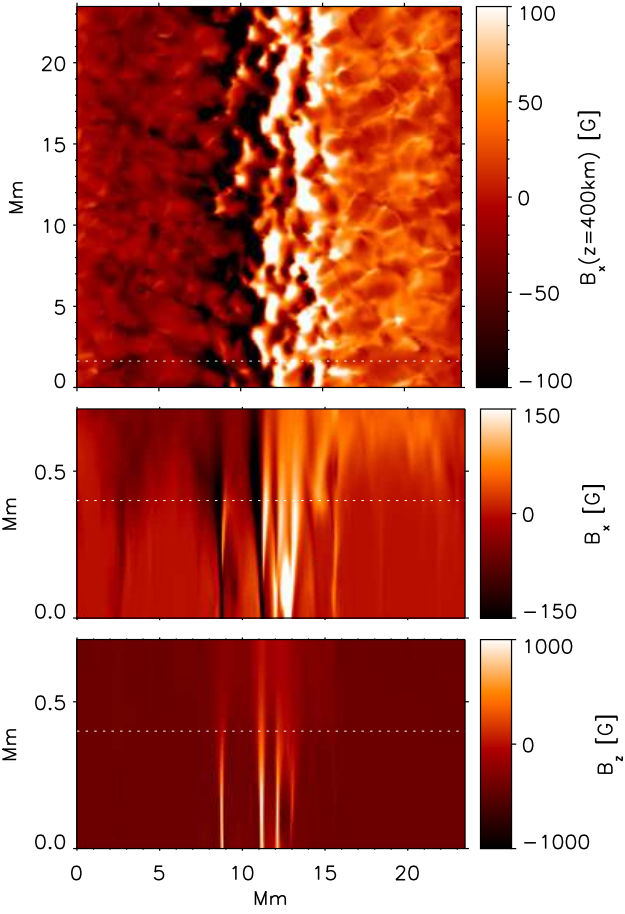}
\caption{As Fig.~\ref{fig:vert-sim} but for the simulation with a potential field at the top boundary. }\label{fig:pot-sim}
\end{figure}

\begin{figure}
\includegraphics[width=10.8cm]{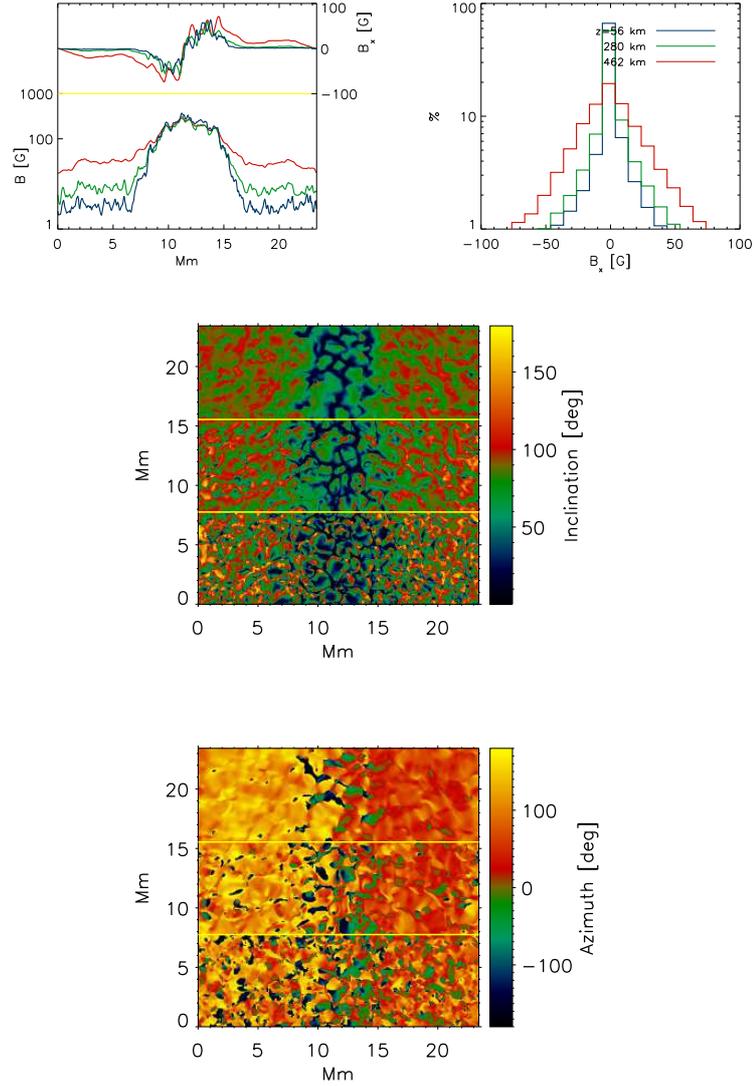}
\caption{Simulation with a potential field at the top boundary. Top left panel: average (over y-direction) x-component of magnetic field (top) and average magnetic field strength (2 lower panels) at three different heights. Top right: histograms of $B_x$ at three different heights. Color code for top panels is given by the legend. Middle row: horizontal cuts of magnetic field inclination at three different heights. Shown is 1/3rd of the domain size for each height. Bottom: as middle row but for the magnetic field azimuth. }\label{fig:pot-exp}
\end{figure}

In contrast, in the potential field BC snapshot
(Fig.~\ref{fig:pot-sim}) the magnetic field fills the entire upper
part of the domain, forming a large scale canopy-like
configuration. Individual flux concentrations are still present close
to $\tau$=1, but towards the top they merge into the
canopy-structure. The canopy structure is very inhomogeneous on small scales and the
effect of \add{down flows related to convective overshooting} is clearly seen. Portions of
the canopy are dragged down to the \add{bottom of the photosphere} by the flows: there is a
significant amount of magnetic field, both vertical and horizontal,
in the intergranular lanes at $z$=0 km. A more detailed view of the potential field BC simulation is given in Fig.\ref{fig:pot-exp}. A large portion of the field outside the strip is horizontal even at low heights, but it is not as homogeneous as higher up. The relative (to the vertical component) strength of the horizontal field increases with height as is seen in the cross-cuts of the average total and horizontal magnetic field strengths. This is caused by both the vertical component becoming weaker with height but also by the horizontal component becoming stronger.

\begin{figure}
\includegraphics[width=8.8cm]{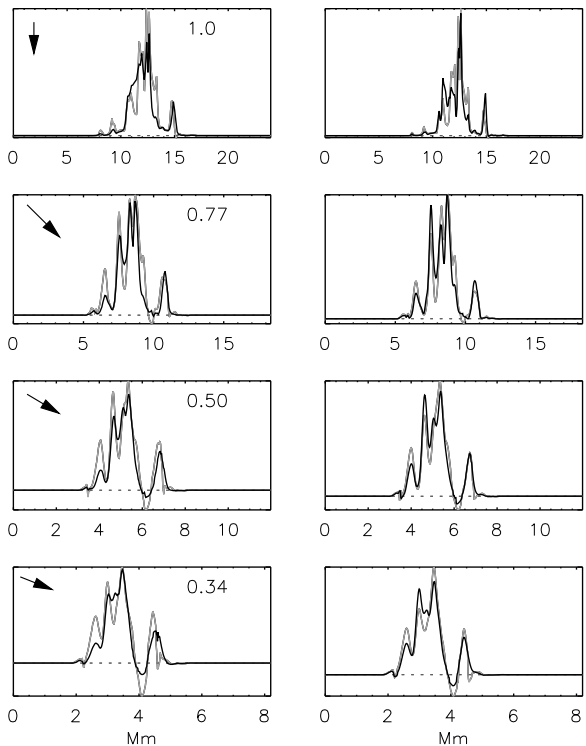}
\caption{Synthetic radial cuts from the vertical field BC simulation at different $\mu$-values (given in the left frames). Left Fe and right NaD$_1$ amplitudes (profiles are not normalized to the continuum intensity). Gray line is at original simulation resolution and black at Hinode resolution. The LOS angle is indicated by the arrow in each figure (on the left). }\label{fig:vert-mur}
\end{figure}

\begin{figure}
\includegraphics[width=8.8cm]{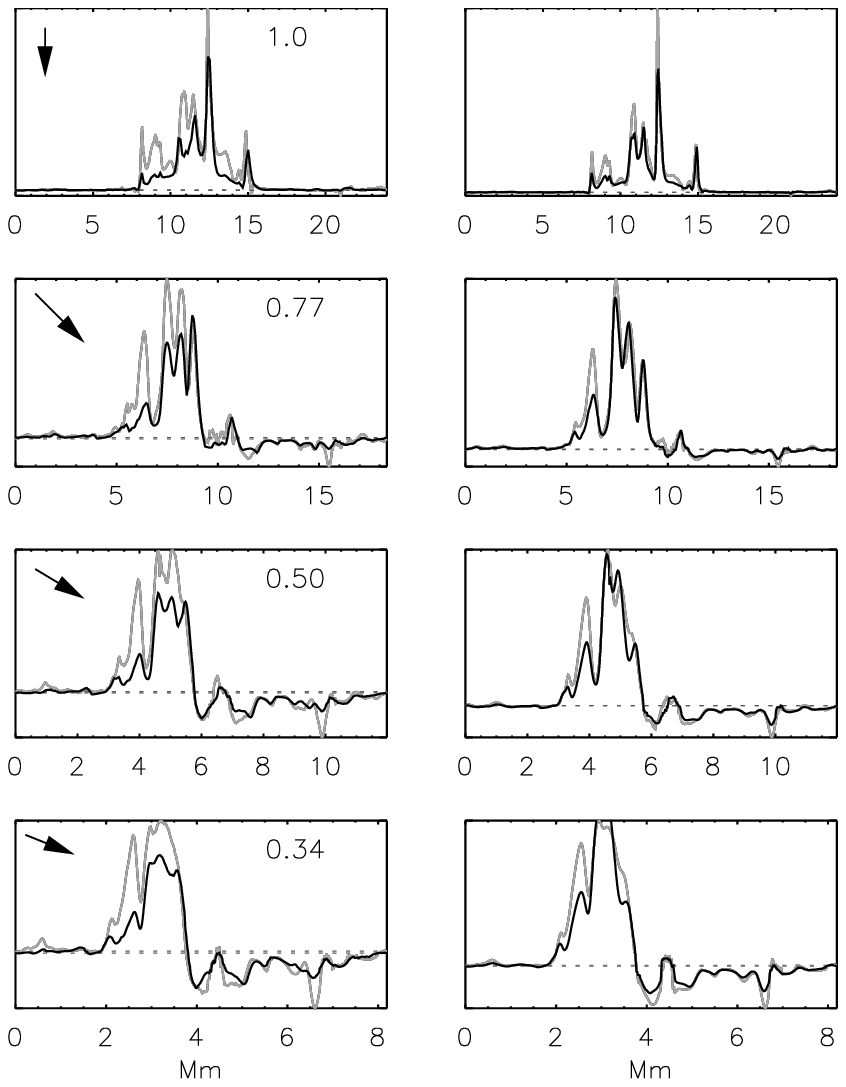}
\caption{As Fig.~\ref{fig:vert-mur} but for the potential field BC simulation. }\label{fig:pot-mur}
\end{figure}

The synthetic observables shown in Figs.~\ref{fig:vert-mur} and ~\ref{fig:pot-mur} give an impression of how observables from
such magnetic configurations would look like. \add{We have not added ordinate scales to the synthetic radial cut figures because the main diagnostic (i.e., the ratio of the diskward to the limbward peak amplitude of Stokes $V$) depends on the expansion and disk position, but not on the amplitude of $V$.} Since the magnetic field
is inhomogeneous at all heights, the resulting cuts exhibit
small-scale structure. Since the spectral lines are formed at a roughly constant optical depth instead of geometric height, the strong changes in density across the simulation domain (flux concentrations and the surroundings) lead to a strongly varying geometric formation height range. The radial cuts do not mimic the cross-cuts of the magnetic field shown in Fig.~\ref{fig:pot-exp}. After taking into account the spectral and spatial resolution of Hinode the cuts from the vertical BC simulation do not at any LOS viewing angle appear bipolar. In contrast, cuts from the potential BC simulation display apparent bipolar features at $\mu \ge $0.34. This is closer to the limb than in the observations or thin flux sheet and tube models ($\mu=0.5$). The ratio decreases with decreasing $\mu$-value. However, the ratios remain fairly large, $\approx$5.

\section{Discussion}

The SOT observations show clearly the expansion of magnetic flux concentrations with height as a change from unipolar radial cuts at disk center to bipolar near the limb. Using the ratio of the radial cuts' center- to the limb-side peak as a proxy for expansion indicates that the magnetic field has fanned out more at the formation height of the NaD$_{1}$ NFI signal than at the formation height of the Fe I 630 nm SP signal. The NaD$_{1}$ ratios are smaller than the Fe I and the switch over from unipolar to bipolar cuts takes place closer to disk center, i.e., at $\mu$=0.5 in NaD$_{1}$ rather than at $\mu \approx$0.3 in Fe I. Furthermore, unlike the SP data the NFI data exhibit exhibit only very few unipolar features close to the solar limb. The apparent expansion between the SP and NFI data and the lack of unipolar features close to the limb in NFI data can be caused by several factors: the NaD$_{1}$ signal is formed higher and the features that would appear unipolar in SP look bipolar in NFI due to increased expansion of the field with height, as expected from flux tube models. Alternatively, the unipolar features are not strong enough or do not reach high enough to be visible in the NFI data. The line formation of the NaD$_{1}$ and Fe I 630 nm lines are somewhat different, e.g., the effective $g$ value of the NaD$_{1}$ line is smaller than for either Fe line making it less sensitive to low magnetic field strengths or fluxes. The difference may be accentuated by other factors such as the NFI data being filtergram while the Fe I signals are deduced from spectrograph data. The different noise levels and pixel sizes of the instruments may also play a role. To fully compare the two one should use simultaneous NFI and SP observations at various disk positions. 

Expansion based on the observed ratios and ratios derived from the thin flux sheet and tube approximations are in good agreement. The synthetic ratios are of similar magnitude and the switch over from unipolar to bipolar takes place at similar $\mu$-values, i.e., around 0.5. Since the sheets expand faster, the thin sheet model results in bipolar features closer to disk center, first measurable ratios appear at $\mu\approx$0.8, and at smaller $r_{0}$: 200 km for the sheet and 400 km for the tube models (assuming the observations are perpendicular to the elongated direction of the sheet).

Unlike in the observations, in the thin sheet/tube models the ratios between the limb-to-center side Stokes $V$ amplitudes and areas of the Fe lines are smaller than the NaD$_{1}$ ratios. While the NaD$_{1}$ line core is formed higher than the Fe line cores, the line flanks, from which the bulk of the Stokes $V$ signal is from, are formed at roughly the same height in the sheet and tube models (based on response functions computed in LTE). The difference between the synthetic and observed ratios is probably at least partly due to the radiative transfer being done in LTE and a sampling bias due to the differences in the SP and NFI data. Also, factors such as model magnetic field strength and microturbulence may influence the ratios. If the disk center-side of the flux concentration has a strong enough field it may become Zeeman saturated before the limb-side, leading to a relatively reduced center side amplitude and a decreased ratio. Since the Fe lines have a larger Zeeman sensitivity this effect would first take place in the Fe lines.

The observations clearly show the expansion of the field which is found to be similar to the expansion derived from using the thin flux sheet and tube approximations. This result is in agreement with \cite{Solanki+others1999}. In other aspects the magnetic network in the Sun is probably not well described by a simple flux tube or sheet scenario. In reality the network is dynamic and likely composed of individual flux concentrations which merge at some height. Observations of chromospheric lines, such as H-$\alpha$ (e.g., \citealt{DePontieu+others2007}), the Ca K (e.g., \citealt{Zirin1974}, \citealt{Pietarila+others2009}) and Ca II infrared triplet lines around 850 nm (e.g., \citealt{Vecchio+others2007}), exhibit fibril structures which probably outline magnetic field lines that are more heated than others, i.e., the fibrillar canopy is thermally inhomogeneous. The magnetic field in contrast is likely quite smooth as indicated by the homogeneous appearance of both the uni- and bipolar patches in the SOT data.

The choice of upper BC for the magnetic field in the MHD-simulations has a significant effect on the expansion properties of the field. The BC affects not only the upper portion of the simulation domain but effects are also seen at the $\tau$=1 level, namely the existence of horizontal and vertical field also outside the initial magnetic strip. In the potential field BC simulations horizontal magnetic field is present everywhere and a canopy-like structure is formed at the top of the domain whereas in the vertical field BC simulation the field remains confined in the strip. \add{We plan to address in a later paper the time evolution of the magnetic flux in the potential field BC simulation to study in detail the transport of magnetic field from the top of the domain, i.e., the canopy, to the bottom of the photosphere.} \add{Note that with the term canopy we are considering the canopy-like expansion of the field, not the actual chromospheric canopy.} The choice of BC affects strongly the synthetic radial cuts: no apparent bipolar features are seen in radial cuts from the vertical BC simulation. In contrast, a bipolar feature begins to appear at $\mu=0.34$ in the potential BC case. The switch over from unipolar to bipolar takes place closer to the limb and the resulting ratios are larger than those seen in the observations or in the zeroth order thin flux tube/sheet models, indicating that the field does not fan out enough.

A stronger expansion in the potential field BC simulation could be achieved by, e.g., having two strips of opposite magnetic polarities in the initial setup or increasing the simulation domain size horizontally. The former would result in a canopy structure connecting the opposite polarities. However, based on the NFI ratios, the magnetic field in the mostly unipolar polar regions expands at a similar rate as in the equatorial limb regions, where mixed polarity fields are found. The latter alternative, larger simulation domain, would result in stronger fanning out because of the periodic BCs in the horizontal direction. Currently the domain size is 24 Mm which is comparable to the typical size of internetwork cells (30 Mm, \citealt{Beckers1968}). Therefore a larger domain would not necessarily be more realistic. Placing the upper boundary higher might also result in stronger expansion \add{since the field would have more volume to expand in}. However, the current treatment of radiative transfer in LTE would become increasingly incorrect. Finally, the field may not yet have reached a potential state at the upper boundary, so that even this boundary condition may not be appropriate.

We plan to extend the current study by using spectropolarimetric observations of photospheric and chromospheric lines to study in detail the network magnetic fields.  

\add{
\subsection{Conclusions}
The circular polarization signal of magnetic flux concentrations is known to change from unipolar at disk center to bipolar near the limb, consistent with the magnetic field fanning out with height. \cite{Solanki+others1999} showed with observations then available that the relative expansion rates of features of various sizes, were consistent
with rates derived for simple thin flux tubes. 
The data from the Hinode satellite, with their high spatial resolution, low noise and freedom from atmospheric seeing effects,
provide a new opportunity to address the question with better statistics and with an emphasis on the variation from disk center to the limb.
We again found the observations to be in good agreement with expansion properties derived from the thin flux sheet and tube approximations,
confirming the previous results with the new data and a more extensive analysis. Since the signal we are interested
in forms higher in the atmosphere near the limb, we found that realistic numerical simulations, which extend only to 700 km
above $log(\tau_{500nm})=0$, are less useful in modeling these observations, with the results strongly depending on the 
choice of boundary conditions for the magnetic field.
}


\acknowledgements{We wish to thank Hector Socas-Navarro for letting us use his Nicole-code. {\it Hinode} is a Japanese mission developed and launched by ISAS/JAXA, with NAOJ as domestic partner and NASA and STFC (UK) as international partners. It is operated by these agencies in co-operation with ESA and NSC (Norway). This work was partly supported by the WCU grant No. R31-10016 from the Korean Ministry of Education, Science and Technology.}

\end{document}